\documentclass[12pt]{iopart}

\usepackage[utf8]{inputenc}
\usepackage[english]{babel}
\usepackage{amssymb}
\expandafter\let\csname equation*\endcsname\relax
\expandafter\let\csname endequation*\endcsname\relax
\usepackage{amsmath}

\usepackage{physics}
\usepackage{perpage}
\usepackage{graphicx}
\usepackage{times}
\MakePerPage{footnote}
\usepackage{underscore}
\usepackage{tikz}
\usepackage{caption}
\usepackage{subfig}
\usepackage{nameref}
\usepackage{braket}
\usepackage{subfiles}
\usepackage{blindtext}
\usepackage{slashed}
\usepackage{multirow}
\usepackage{simplewick}
\usepackage{verbatim}
\usepackage{tikz}
\usepackage{caption}
\usepackage{subfig}
\usepackage{bm}
\usepackage{frontespizio}
\usepackage{algorithmicx}
\usepackage{algorithm}
\usepackage{algpseudocode}

\usepackage{makecell, caption, booktabs, multirow, siunitx}
\usepackage[square, numbers]{natbib}
\usepackage{url}
\usepackage{microtype}

\usepackage{xr}
\usepackage{hyperref}

\usepackage{afterpage}

\begin{document}
\title{Solving Rubik's Cube via Quantum Mechanics and Deep Reinforcement Learning}
\author{Sebastiano Corli}
\address{Istituto di Fotonica e Nanotecnologie, Consiglio Nazionale delle Ricerche, Piazza Leonardo da Vinci 32, I-20133, Milano, Italy}
\address{Dipartimento di Fisica Aldo Pontremoli, Università degli Studi di Milano, via Celoria 16, I-20133 Milano, Italy}
\author{Lorenzo Moro}
\address{Istituto di Fotonica e Nanotecnologie, Consiglio Nazionale delle Ricerche, Piazza Leonardo da Vinci 32, I-20133, Milano, Italy}
\address{Politecnico di Milano, Via Colombo 81, I-20133, Milano, Italy}
\author{Davide E. Galli}
\address{Dipartimento di Fisica Aldo Pontremoli, Università degli Studi di Milano, via Celoria 16, I-20133 Milano, Italy}
\author{Enrico Prati}
\address{Istituto di Fotonica e Nanotecnologie, Consiglio Nazionale delle Ricerche, Piazza Leonardo da Vinci 32, I-20133, Milano, Italy}

\ead{enrico.prati@cnr.it}
\vspace{10pt}
\begin{indented}
\item[]August 2021
\end{indented}
\maketitle

Rubik's Cube is one of the most famous combinatorial puzzles involving nearly $4.3 \times 10^{19}$ possible configurations. Its mathematical description is expressed by the Rubik's group, whose elements define how its layers rotate. We develop a unitary representation of such group and a quantum formalism to describe the Cube from its geometrical constraints. Cubies are described by single particle states which turn out to behave like bosons for corners and fermions for edges, respectively. 

When in its solved configuration, the Cube, as a geometrical object, shows symmetries which are broken when driven away from this configuration. For each of such symmetries, we build a Hamiltonian operator. When a Hamiltonian lies in its ground state, the respective symmetry of the Cube is preserved. 

When all such symmetries are preserved, the configuration of the Cube matches the solution of the game. To reach the ground state of all the Hamiltonian operators, we make use of a Deep Reinforcement Learning algorithm based on a Hamiltonian reward. The Cube is solved in four phases, all based on a respective Hamiltonian reward based on its spectrum, inspired by the Ising model.

Embedding combinatorial problems into the quantum mechanics formalism suggests new possible algorithms and future implementations on quantum hardware. 


\section{Introduction}

Combinatorial optimization includes a vast class of problems, ranging from the traveling salesman problem~\cite{travelling2,travsalesman} and protein folding~\cite{cinesi} to the minimum spanning tree~\cite{spanning}.
As the lack of a metric forbids the measurement of the distance to the goal configuration, methods such as Genetic algorithms~\cite{travelling1,travelling2,travelling3,travelling4} or Monte Carlo Tree Search have been employed~\cite{MCTS}. Recently, such method has been combined with a Reinforcement Learning (RL) algorithm, to develop a solver for the Rubik's Cube~\cite{california}. The class of RL algorithms is based on training an agent (a software) to accomplish a specific task, for example solving the Rubik's Cube.
In statistical mechanics, the Cube has some analogies with systems in which energy minimization processes are involved, such as the protein folding~\cite{cinesi,california}. In such systems, many possible states are available, but the stable (or native) configuration is unique. In fact, the phase space of the Rubik's Cube is huge, as there are approximately $4.3 \times 10^{19}$ available configurations~\cite{janetchen}, while only one configuration corresponds to the solution. 
The Rubik's Cube has been explored in the past for its group properties~\cite{gruppilibro,20mosse,janetchen,NP,NP2}. It has been shown~\cite{NP,NP2} that, generalizing the $3 \times 3 \times 3$ Cube to a $n\times n \times n$, such combinatorial problem scales as a NP-complete one. Many NP-complete problems can be mapped into a spin formulation~\cite{isingNP}. Ising model has been exported in other fields~\cite{isingsocial,isingeconomic,rocutto2020quantum} and suits with optimization problems, especially for discrete systems~\cite{IsingGS,IsingGS2}. Furthermore, Ising model is naturally embedded on the core of adiabatic computers such as D-Wave~\cite{DWave}.

Reinforcement Learning has been applied to solve many competitive games (chess for example) without previous human knowledge~\cite{tesauro,human,mastering,silver2017mastering2}, but also to protein folding problems~\cite{alquraishi2019alphafold}. In the field of quantum mechanics, Reinforcement Learning has been employed for Molecular Design~\cite{molecules}, or to find the ground state of a Hamiltonian as an optimal control problem~\cite{optimalControl}. New optimisation techniques for arbitrary Hamiltonians are rising exploiting the calculus power of quantum computers themselves~\cite{IBMoptimisation}. Reinforcement Learning has quite recently been proposed for the control of quantum systems~\cite{Porotti1, Porotti2, Porotti3, Porotti4, Porotti5, Porotti6, Porotti7, Porotti8}, along with a strictly quantum Reinforcement Learning implementation~\cite{Porotti6, Porotti9}.
%
In the past, we have already applied Deep Reinforcement Learning (DRL) algorithms to quantum problems, including coherent transport of quantum states~\cite{porotti2019coherent} and digitally stimulated Raman passage~\cite{paparelle2020digitally}. Furthermore, we have applied an unsupervised learning algorithm on a quantum hardware, the D-Wave annealer, expressed as a QUBO problem~\cite{rocutto2020quantum}. Reinforcement Learning has been applied also in the field of quantum compiling, to apply a set of logic gates~\cite{maronese2021continuous} on a quantum hardware~\cite{moro2021quantum}.
%
Instead, here we employ quantum mechanics to set an algebraic environment, where to apply a DRL algorithm to solve the Rubik's cube as a quantum Hamiltonian minimization problem. Remarkably, we exploit only physical rewards, namely the reward functions used to return the feedback to the DRL agent, are based on the Ising model. In principle it would enable to implement such formalism directly on a quantum hardware like the above mentioned D-Wave, to solve the Rubik's Cube and similar combinatorial problems.
We start from Chen's formulation of the Rubik's group~\cite{janetchen}, by translating its geometrical postulates in the group action on a vector in a Hilbert space.
The configurations of the Cube (used as the inputs $s$ for the DRL agent) are represented by vectors $\ket{s}$ in a Hilbert space, while each element of the Rubik's group (the outputs $a$, which correspond to the actions returned by the DRL agent) is a unitary transformation $U$. It is then possible to build a distance in the phase space of the Cube via Hamiltonian operators $\hat{H}$, so that the energy observable $E$ can be used to evaluate the distance between two states. The solved state coincides with the ground state. In this framework, the energy levels map a distance from the solved configuration to any other. Ideally, the further the system is from the solved state, the higher the Hamiltonian eigenvalues should be. However, this condition could not be held, as far as the Hamiltonians are used to define a distance between two states, but they are not the real distance in the phase space. For the sake of brevity, in the following we refer to this method of defining a distance between states by using the energy spectrum as energy metric. The key point is to build good Hamiltonian operators, so that when the system is driven away from the solution, higher values in the Hamiltonian spectrum should raise negative feedback to the agent. Such feedback, which is called reward in RL, is the key to make the agent learning and solving the problems. The core of the algorithm we built is to set the reward returned to the agent as a function of the Hamiltonian spectrum of the Cube.

To solve the Cube, we split the process into four phases. Each phase is equipped with its own Hamiltonian, so that the overall energy metric of the Cube is mapped by four operators, a far easier task than using a unique metric for $4.3 \times 10^{19}$ configurations. The aim of each phase is to reach the ground state for an appropriate Hamiltonian, and a set of moves able to accomplish this purpose. One key point, for each phase, consist of building such Hamiltonian operator and a suitable set of moves from the Rubik's group, such as the twelve rotations of the Cube faces or some compositions. To describe the Cube as a quantum system, it has to be endowed with some physical quantities first, such as momentum and spin, to form a complete set of observables. In this way, we can identify the solved state from any other, and build the four Hamiltonian operators. To build the Hamiltonians, we took inspirations from the Ising model, and the interaction between spins.


The development of a solver for the Cube via a quantum formalism is done by the unitary representation for the Rubik's group, whose purpose is accomplished in the first section. In the second section, we show how a problem of combinatorial optimization can be translated into an energy minimization in the Rubik's cube, treated as an Ising system, and how the Hamiltonian spectrum can be employed in the reward for an agent in a Reinforcement Learning algorithm, so to achieve the solution of the Cube. To reach the solution, quantum phenomena such as tunneling, superposition or entanglement will not be exploited, the algorithm has been thought to be a classical solver. In order to demonstrate the validity of the unitary formalism, we show how the Cube can be solved within this formalism in a classical fashion, while further improvements and methods, for example developing a quantum solver on a quantum hardware, are left to future works. The work is organized as follows: Section $\ref{sec:CubConf}$ recaps the geometry of the Rubik's Cube, Section $\ref{sec:QuantumFormalism}$ is aimed to provide the new quantum formalism to describe the Cube, Section $\ref{sec:results}$ features the strategy developed to solve the Cube by this formalism and a DRL algorithm, and the last Section $\ref{sec:Conclusions}$ draws the conclusions.

\section{Definitions and formalism}
\label{sec:CubConf}

\begin{figure}
\subfloat{\includegraphics[width=.65\textwidth]{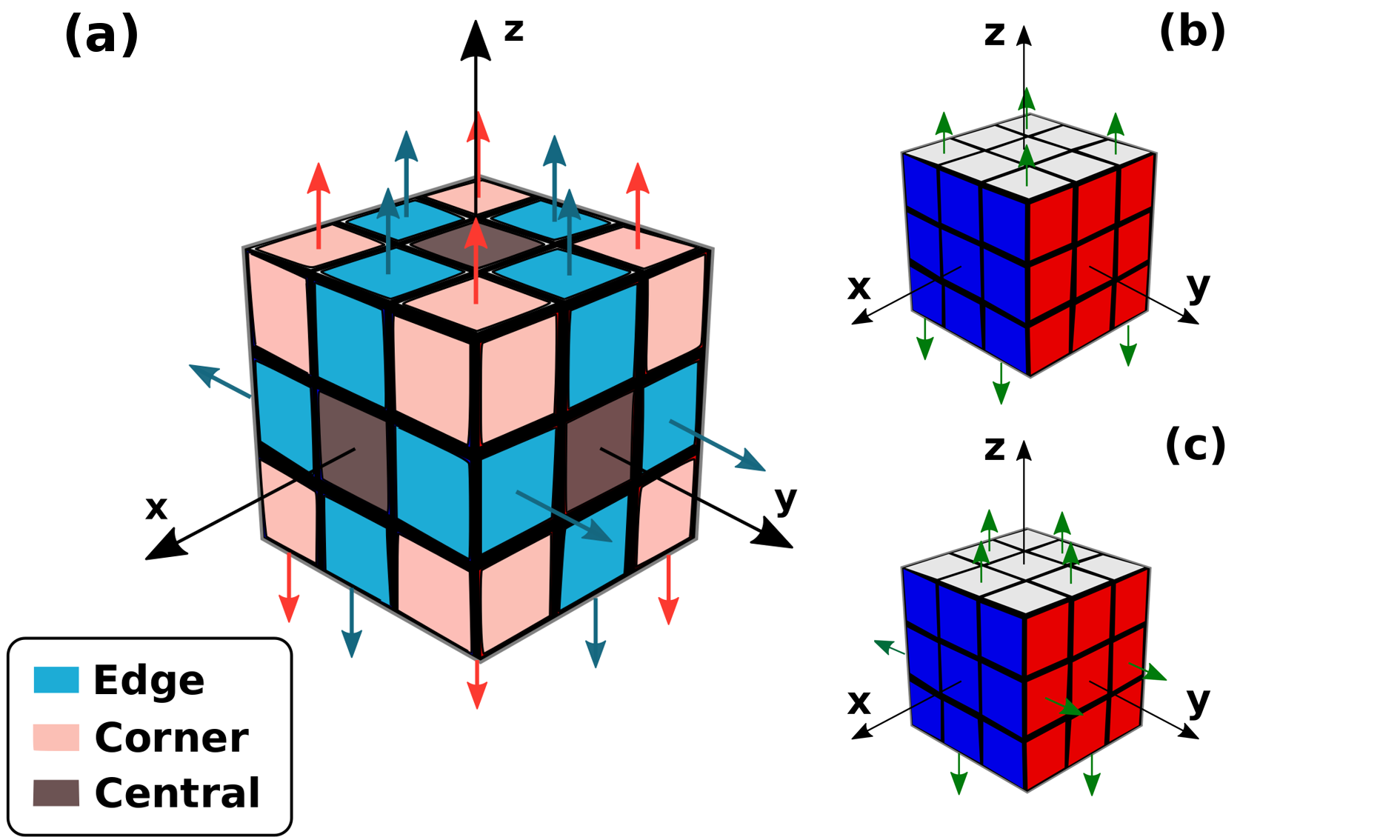}}\\
\centering\caption{(a) the different kind of cubies composing the Cube: in light blue, in pink and in grey the edges, the corners and the centrals respectively. The arrows define the arbitrary choice of orientation for the solved state. Two different colours are applied for corners (pink) and edges (blue). (b) The faces of the Cube with their conventional colours, blue, red and white. Coordinated axes are introduced to describe the positions of the cubies, and the choice for the orientation in the solved configuration for the sole corners. (c) The same coordinated axes to describe the position of the cubies and the choice of orientation for the edges in the solved configuration.}
\label{fig:start}

\end{figure}

The Cube is composed by $26$ smaller cubes, called cubies~\cite{janetchen}. Six of them show one face (the centrals), which cannot be moved from their relative positions, $12$ edges show two faces and $8$ corners show three faces (the corners). A visual description is shown in Figure $\ref{fig:start}$.
A corner cubie cannot take the place of an edge one, and vice versa. Any configuration of the Cube can be univocally identified by two sets of features, namely the positions and orientations of the cubies. The position of a cubie marks how far a cubie is from its place in the solved configuration. The orientation of a cubie stores how, keeping the cubie in its solved position, it has been rotated around a rotation axis suitable to induce permutations. Orientation is graphically represented as an arrow on a face of the cubie, as in Figure $\ref{fig:start}$.
Orientations and positions are modified by rotating the layers of the Cube. These transformations are elements from Rubik's group $(\mathfrak{R}, \circ)$, which consists of six generators, each of them corresponding to the rotation of a coloured face. In Singmaster notation~\cite{singmaster}, they are labelled as

\begin{equation} \label{eq:Rubikgroup}
 \mathfrak{R} = \{U, D, F, B, L, R\}
\end{equation}

where $U$, $D$ are the rotations of the upper and lower faces, $F$, $B$ front, back, $L$, $R$ left, right respectively.
To describe the orientation and position of each cubie, one may set an orientation for the solved configuration and three cartesian axes (Figure $\ref{fig:start}$). The geometry of orientation is affected by the number of faces displayed by the cubie. The ways in which a cubie can be oriented are the way the same cubie can be put in a fixed position by rotating its own faces. Any allowed configuration can be mapped from the solved one, once an algebraic representation for the Rubik's group is given. 
The six cubies fixed in the middle of the faces are called \textit{centrals}. They are invariant under all the transformations and show just one face, thus they are not associated to any specific orientation.
\textit{Edge} cubies show two faces, thus two possible modes of orientation are available. Two orientation are available, the orientation in the solved configuration and its flipped arrangement of the two faces. The orientation we choose for the edges is shown in Figure \ref{fig:start}(c).
\textit{Corner} cubies exhibit three faces, which can be arranged in three different manners. Swapping two corner faces is not allowed by the Rubik's group~\cite{janetchen}. The three faces of a corner cubie can be rotated clockwise or anti-clockwise, allowing three possible orientation. The orientation we choose for the corners is shown in Figure \ref{fig:start}(b).

\section{Quantum Formalism}
\label{sec:QuantumFormalism}

In order to introduce a Hamiltonian approach to a Reinforcement Learning problem (as explained in Section $\ref{sec:results}$), we provide a description of the Cube based on a quantum formalism. The core of this quantization lies in developing a unitary representation of the Rubik's group. By introducing a complete set of observables, and studying how any transformation from Rubik's group affects these observables, it is possible to introduce a unique unitary representation for it.\\
Corners and edges own different transformation properties so they are treated separately.


\subsection{Single particle state}

Two geometrical properties identify uniquely the state of a cubie, as clarified in Section $\ref{sec:CubConf}$, namely its position and its orientation. The state of position and orientation can be labelled by four numbers consisting of a triple $k_i$ ($i=1,...,3$ for each axis) and $s$. A description of such a state is given by a vector in a Hilbert space $\mathcal{H}$:
\begin{equation} \label{eq:s&k}
    \ket{cubie} = \ket{\textbf{k},s} \in \mathcal{H}
\end{equation}
The focus is now to build a representation, i.e. introducing four operators such that
\begin{equation}\label{eq:eigenvalues}
    \hat{K_i} \ket{\textbf{k},s} = k_i \ket{\textbf{k},s}, \; \hat{S} \ket{\textbf{k},s} = s \ket{\textbf{k},s}
\end{equation}
$\hat{K}_i$ ($i=1,2,3$) and $\hat{S}$ form a complete set of observables, i.e. their eigenvalues completely define the system without degeneracy. The statement of completeness requires $\hat{K}_i$ and $\hat{S}$ to have $\ket{\textbf{k},s}$ as simultaneous eigenvector and thus to commute:

\begin{equation} \label{eq:commuting_complete}
    [\hat{K}_i; \hat{S}]=0, \quad [\hat{K}_i; \hat{K}j]=0
\end{equation}
The absence of a real dynamics, and thus the lack of a Hamiltonian, prevents to define the explicit expression of $\ket{\textbf{k},s}$ via Schr{\"o}dinger or matrix equations. To accomplish this aim, such a state must be built by studying the action of the transformations from the Rubik's group, $\mathfrak{R}$, on it.

\subsection{A description for the position of the cubies}
\label{sec:position}

\begin{figure}
\quad
\subfloat[][]{\includegraphics[width=.35\textwidth]{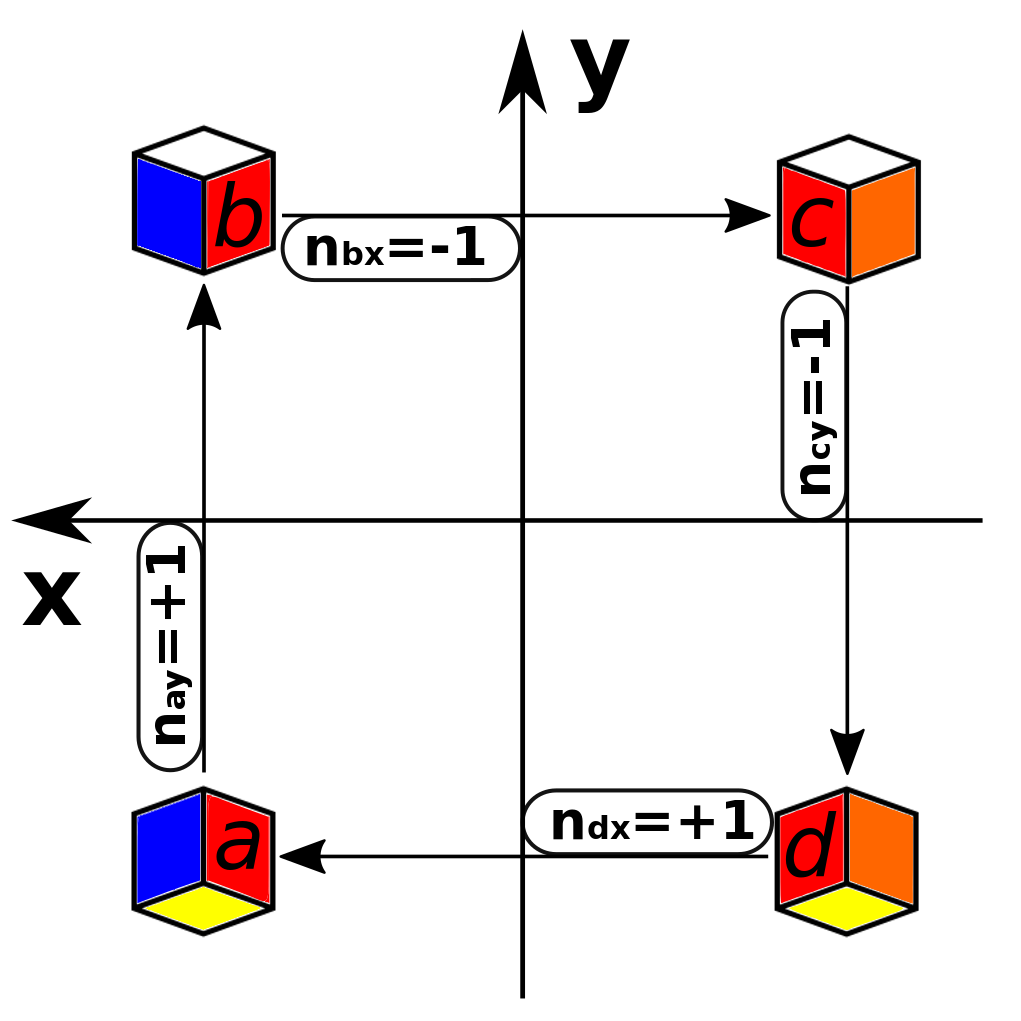}}
\subfloat[][]{\includegraphics[width=.35\textwidth]{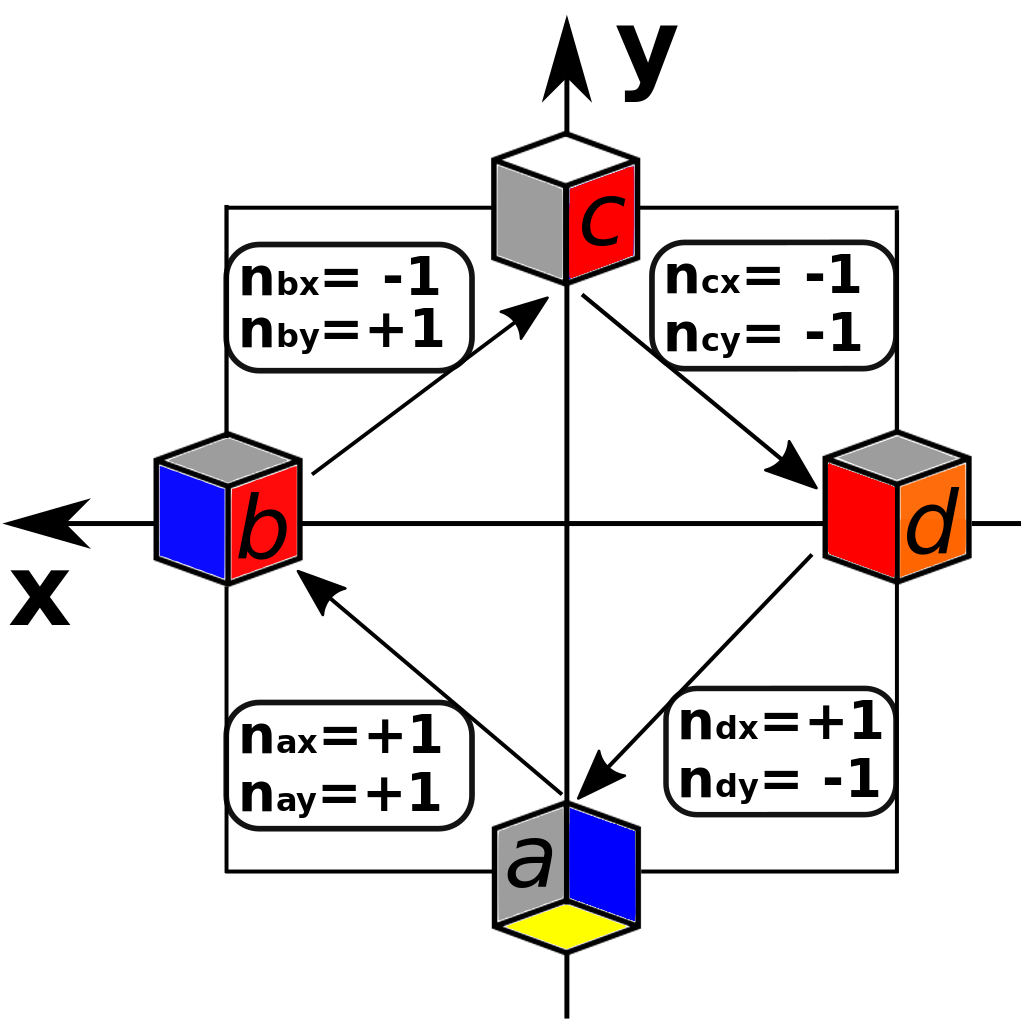}}
\centering\caption{Schematic representation of a F transformation, i.e. a rotation of the frontal (red) layer as from Figure $\ref{fig:start}$. The position of each cubie can be described by three quantum numbers, $n_x$ , $n_y$ and $n_z$, each of them is set to be zero in the solved configuration of the Cube. (a) Effect of the $F$ rotation (rotation of the frontal red layer of Figure $\ref{fig:start}$) on the corners by a translation for each corners. The \textit{d} cubie is moved along the x-axis of one position from its solved position, which is labelled by raising the quantum number $n_{dx}$ by one ($n_x$ is now $+1$). Conversely, \textit{c} is translated in the opposite direction along the y-axis by one position, which means its quantum number $n_{cy}$ will be lowered by one ($n_y$ is now $-1$). (b) Effect of the $F$ rotation by pairs of translations for each edge. The edges are moved diagonally, which makes the cubies to shift along both x and y directions. Thus, for each cubie, both $n_x$ and $n_y$ will be raised or lowered in a $F$ transformation.}
\label{fig:positions}
\end{figure}
Any transformation from $\mathfrak{R}$ affects the position of the involved cubies. By rotating a layer of the Rubik’s Cube each cubie shifts along a particular axis. As far as any rotation is by $\pi/2$, the symmetry of translation is discrete. Similar symmetries recur in crystalline structures for instance, where each atom is set on a particular site over a lattice. A translation operator $\hat{U}(\textbf{T})$ over $\textbf{k}$ should accomplish the following property:
\begin{equation}
    \hat{U}(\textbf{T})\ket{\textbf{k},s}=\ket{\textbf{k}+\textbf{T},s}, \quad \hat{U}(\textbf{T}) : \mathcal{H} \to \mathcal{H}
\end{equation}
The following properties must hold:
\begin{enumerate}
\item the operator of translation must be unitary, i.e.
\begin{equation}
    \hat{U}(\textbf{0})=\hat{U}(\textbf{T})\hat{U}(-\textbf{T})=\hat{U}(\textbf{T})\hat{U}^\dagger(\textbf{T})=\hat{U}(\textbf{T})\hat{U}^{-1}(\textbf{T})=\hat{\mathbb{I}}
\end{equation}
\item any $\hat{U}(\textbf{T})$ translation must be discrete. This condition can be written as $\textbf{T}=a\textbf{n}$, $a$ being a real scalar number and $n \in \mathbb{Z}$;
\item it must also be periodic to preserve the symmetry of the Cube;
\end{enumerate}
An explicit representation for $\ket{\textbf{k},s}$ which respects the conditions above is given by the wave function for a free particle with periodic boundary conditions:
\begin{equation}
    \psi_\textbf{n}(\textbf{x})=\frac{1}{\sqrt{V}}e^{i\textbf{k}\cdot\textbf{x}}
\end{equation}
where $\textbf{k}=2\pi \textbf{n}/l$ and $l^3=V$. Such a wave function is defined in a $L^2[0;l]$ space, endowed with a scalar product and thus a norm (as any Hilbert space):
\begin{equation}
    \braket{\psi_\textbf{n}|\psi_\textbf{n}} = \frac{1}{V} \int_V \dd^3x \; e^{-i\textbf{k}\cdot\textbf{x}}e^{i\textbf{k}\cdot\textbf{x}} = 1
\end{equation}
To realize the eigenvalues equation (\ref{eq:eigenvalues}), the $\hat{K}_i$ operators act on such a state as a derivative:
\begin{equation}
    \hat{K}_i \ket{\textbf{k},s}=-i\frac{\partial}{\partial x_i} \psi_\textbf{n}(\textbf{x})\ket{s}=k_i\psi_\textbf{n}(\textbf{x})\ket{s}=k_i\ket{\textbf{k},s}
\end{equation}
Because $k_i$ and $x_i$ are conjugated variables, the operator of translation can be written as
\begin{equation}
    \hat{U}(\textbf{T}) = e^{i\textbf{T} \cdot \textbf{x}}, \quad \hat{U} : L^2[0;l] \rightarrow L^2[0;l]
\end{equation}
which satisfy the request to be unitary. As $\textbf{k}$ must be proportional to a vector $\textbf{n}$ of integers, any translation by $\textbf{T}$ must preserve such proportionality:
\begin{equation}
    \textbf{k'} = \textbf{k}+\textbf{T}\equiv \frac{2\pi}{l} \textbf{n'}
\end{equation}
which implies $\textbf{T}=2\pi\textbf{t}/l$, for $\textbf{t}=(t_1,t_2,t_3)$, $t_i \in \mathbb{Z} \, \forall i$. Thus the second condition is satisfied.
The variable $\textbf{x}$ does not denote any physical observable, it is just the generator of $\textbf{T}$ translations.\\
The operator $\hat{\textbf{K}}$ marks the position in the reciprocal space. Each cubie, when the Cube lies in its solved state, is defined by $\textbf{k}=\textbf{0}$. To account for the distance of a cubie from its solved position, we rely on the $\hat{\textbf{N}}$ operator, defined as
\begin{equation} \label{eq:Naction}
    \hat{N}_i \psi_\textbf{n}(\textbf{x}) = \frac{l}{2\pi}\hat{K}_i \psi_\textbf{n}(\textbf{x}) = -i \frac{l}{2\pi} \frac{\partial}{\partial x_i} \psi_\textbf{n}(\textbf{x}) = n_i \,\psi_\textbf{n}(\textbf{x})
\end{equation}
$\textbf{n}=(n_x,n_y,n_z)$ is the number of steps along the grid that the cubie has taken from its solved position along $x$,$y$ and $z$ axes. This vector $\textbf{n}=(n_x,n_y,n_z)$ defines univocally the position of each cubie, with respect to their allocation in the solved configuration of the Cube, for which $\textbf{n}=(0,0,0)$ for all the cubies.

\subsection{A description for the orientation of the cubies}
\label{sec:orientation}

Considering any edge in Figure $\ref{fig:start}$, its faces can be disposed in the state shown in the figure or in the reverse one, obtained by flipping them. A vectorial description for $\ket{s}$ edge cubies can be set as 
\begin{equation} \label{eq:edges}
\ket{\uparrow} = \begin{pmatrix} 1 \\ 0 
\end{pmatrix}, \; \ket{\downarrow} = \begin{pmatrix}  0 \\ 1
\end{pmatrix},
\end{equation}
To flip such a dichotomous state, we introduce the operator $\sigma_x$ and another observable $\sigma_z$ to measure the state of orientation:
\begin{equation} \label{eq:pauli}
\sigma_x = \begin{pmatrix} 0 & 1 \\ 1 & 0
\end{pmatrix}, \quad
\sigma_z^e = \begin{pmatrix} 1 & 0 \\ 0 & -1
\end{pmatrix}
\end{equation}
where $\sigma_x$ and $\sigma^e_z$ are the first and the third generators~\cite{sakurai,picasso} for $\mathfrak{su}(2)$ algebra. The $e$ symbol on $\sigma_z^e$ refers to the fact this matrix deals with edge states.

Referring to Figure $\ref{fig:start}$, any corner may lie in the actual state or may be rotated clockwise or anti-clockwise. The corner orientation takes three possible states into account:
\begin{equation} \label{eq:corners}
\ket{+} = \begin{pmatrix} 1 \\ 0 \\ 0
\end{pmatrix}, \; \ket{0} = \begin{pmatrix} 0 \\ 1 \\ 0
\end{pmatrix}, \; \ket{-}= \begin{pmatrix} 0 \\ 0 \\ 1
\end{pmatrix},
\end{equation}{}
Operators for clockwise, anti-clockwise rotations on the states above and a corresponding observable are given by
\begin{equation} \label{eq:pauli2}
\sigma_{x_A} = \begin{pmatrix}
0 & 1 & 0 \\
0 & 0 & 1 \\
1 & 0 & 0 \\
\end{pmatrix},  \; \sigma_{x_C} = \begin{pmatrix}
0 & 0 & 1 \\
1 & 0 & 0 \\
0 & 1 & 0 \\
\end{pmatrix}, \quad
\sigma^c_{z} = \begin{pmatrix}
1 & 0 & 0 \\
0 & 0 & 0 \\
0 & 0 & -1 \\
\end{pmatrix}
\end{equation}
where $A$ and $C$ indices refer to clockwise and anti-clockwise, while the $c$ symbol on $\sigma^c_z$ refers to the fact this matrix deals with corner states. It is straightforward to check the following properties:
\begin{equation} \label{eq:proppauli2}
[\sigma_{x_A}]^3 = [\sigma_{x_C}]^3 = \mathbb{I}, \qquad \sigma_{x_A} \sigma_{x_C} = \mathbb{I}, \qquad \sigma_{x_C}^T = \sigma_{x_A}
\end{equation}
This representation recalls the spin formalism, where corners are boson-like quantum objects with total momentum $S^2=1$ and edges fermion-like quantum objects with $S^2={1}/{2}$. The transformations from Rubik's group either may or may not modify the state of orientation of the cubies, depending on the choice of orientation made in the solved configuration. For example, in Figure $\ref{fig:start}$, where the solved state of the Cube is shown, an upper rotation does not affect the global (neither the local) state of orientation, as the arrows are disposed in the same way after an $U$ (up) transformation. The Cube is correctly oriented when all the cubies are.

The norm of the state $\ket{s}$, for both corners and edges, descends from the euclidean metric in a vector space. Being the vectors in equation ($\ref{eq:edges}$) the generators for an orthonormal basis in $\mathbb{R}^2$, and the vectors in equation ($\ref{eq:corners}$) for $\mathbb{R}^3$, the scalar product can be induced to be
\begin{equation}
    \braket{s=\alpha|s=\beta} = \delta_{\alpha \beta}
\end{equation}


\subsection{Multiparticle state of the Cube}

Combining the results from Section $\ref{sec:position}$ and $\ref{sec:orientation}$, the state of a single cubie $\ket{\textbf{k}, s}$ belongs to  $\mathcal{H}_e=L^2[0;l]\otimes\mathbb{R}^2$ space for edges, and $\mathcal{H}_c=L^2[0;l]\otimes\mathbb{R}^3$ for corners, respectively. As the tensor product of two Hilbert spaces is a Hilbert space, the corner and edge vector spaces are Hilbertian.
The Rubik's Cube is built by wedging twenty cubies together, i.e. its global state is described  by $20$ single particle states. The global state $\ket{C}$ can then in turn be described as a few-bodies system. Introducing a new observable $\hat{S}^2$ in the single particle states of each cubie, it is possible to distinguish corners (bosons) from edges (fermions):
\begin{equation}
    \ket{C} = \bigotimes_{i=1}^{12} \psi_{\textbf{n}_i}(\textbf{x}_i)\ket{s_i^2=\frac{1}{2},s_i}\otimes \bigotimes_{i=13}^{20} \psi_{\textbf{n}_i}(\textbf{x}_i)\ket{s_i^2=1,s_i}
\end{equation}
where $s_i^2$ is the quantum number of $\hat{S}_i^2$ for each $i$-th cubie, $i$ is an index referring to each cubie. In the first tensor product, $i$ runs from $1$ to $12$ because there are twelve edges in the Cube, in the second from $13$ to $20$ because there are eight corners.
In the spinor formalism, when the Cube is oriented, such tensor product is given by
\begin{equation}
    \ket{C} = \begin{pmatrix} \psi_{\textbf{n}_1}(\textbf{x}_1) \\ 0 
\end{pmatrix} \otimes ... \otimes 
\begin{pmatrix} \psi_{\textbf{n}_{12}}(\textbf{x}_{12}) \\ 0 
\end{pmatrix} \otimes \begin{pmatrix} 0 \\ \psi_{\textbf{n}_{13}}(\textbf{x}_{13}) \\ 0 
\end{pmatrix} \otimes
... \otimes
\begin{pmatrix} 0 \\ \psi_{\textbf{n}_{20}}(\textbf{x}_{20}) \\ 0 
\end{pmatrix} 
\end{equation}
Moreover, if the Cube is solved, $\textbf{n}_i=(0,0,0) \, \forall i=1,...,20$, or rather
\begin{equation}
    \hat{N}^2\ket{C}=0
\end{equation}
where $\hat{N}^2=\sum_{i=1}^{20} \hat{\textbf{N}}_i^2$, $i$ being the index for the cubies and $\hat{\textbf{N}}$ the operator defined in equation ($\ref{eq:Naction}$), and $\hat{\textbf{N}}^2=\hat{N}_x^2+\hat{N}_y^2+\hat{N}_z^2$.

\begin{figure}
\subfloat{\includegraphics[width=.50\textwidth]{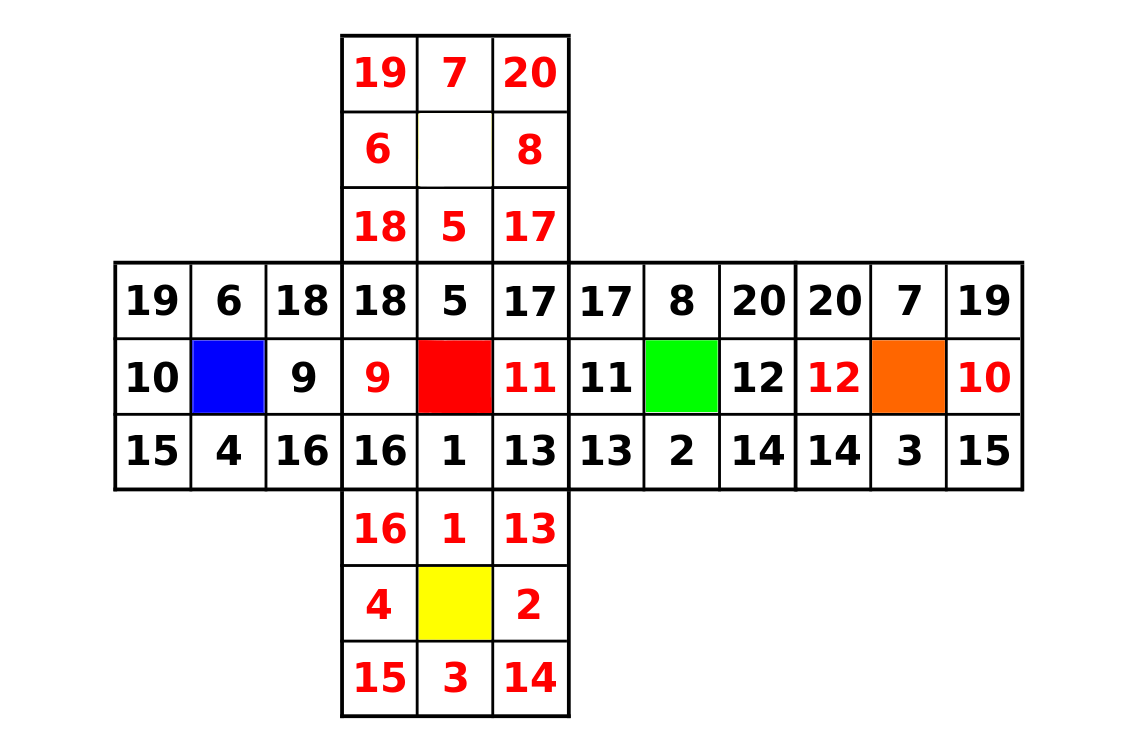}}
\centering\caption{The Rubik's Cube, in the solved configuration, opened to show all of its faces. The cubies are enumerated in order to describe the transformations of the Rubik's group. The red numbers show the oriented faces of the cubies, i.e. the arrows in Figure $\ref{fig:start}$. The colours at the center of each face are those of the Cube: red stands for the front face F, blue for the left F, green for the right R, orange for the back B, yellow for the bottom U, white for the top D. In Figure $\ref{fig:positions}$, the $a$, $b$, $c$, $d$ edges are respectively the cubies $1$, $9$, $5$ and $11$ in the figure above, while the $a$, $b$, $c$, $d$ corners the cubies $16$, $18$, $17$, $13$.}
\label{fig:grid}
\end{figure}

\subsection{Action of the Rubik's group}
\label{sec:action}

Corners and edges cannot be transformed into each other. This constraint imposes the transformations from equation ($\ref{eq:Rubikgroup}$) to act separately on corners and edges, so a $U$ transformation can be built as
\begin{equation}
    \hat{U}=\hat{U}_e\otimes \hat{U}_c
\end{equation}
Every element from Rubik's group performs eight translations, four on edges and four on corners, as shown in Figure $\ref{fig:positions}$. In such figure, $a$ corner is translated along $y$ by one step, which means $n_y^a \rightarrow n_y^a+1$, while the $a$ edge in a diagonal way, such that $n_y^a\rightarrow n_y^a+1$ and $n_x^a\rightarrow n_x^a-1$. A unitary representation of such translations is

\begin{equation}
    \hat{U}_e(\textbf{T}_a) = e^{i\frac{2\pi}{l}(-x_a+y_a)}, \quad \hat{U}_c(\textbf{T}_a) = e^{i\frac{2\pi}{l}y_a}
\end{equation}
After performing a translation over the momenta $\textbf{k}$ (or equivalently, the steps \textbf{n}), the involved cubies are permuted, following the direction of the arrows as shown in Figure $\ref{fig:positions}$. An operator $\hat{P}_{\sigma(a,b,c,d)}$ of permutations is introduced in the overall transformation, whose action is given by
\begin{equation}
    \hat{P}_{\sigma} : \bigotimes_{i=1}^{20} \mathcal{H}_i \to \bigotimes_{\sigma(i)=1}^{20} \mathcal{H}_{\sigma(i)}, \quad \hat{P}_{\sigma}\ket{1}\ket{2}...\ket{20}=\ket{\sigma(1)}\ket{\sigma(2)}...\ket{\sigma(20)}
\end{equation}
It holds from Figure $\ref{fig:positions}$ that $\sigma(a,b,c,d)=(d,a,b,c)$ for both corners and edges, which means $d\rightarrow a$, $a\rightarrow b$ and so on.



The transformations from $\mathfrak{R}$ can also affect the state of orientation of the Cube (not all of them, as we have seen from Section $\ref{sec:orientation}$). When acting on the orientation of the cubies, the edge transformation $\hat{U}_e$ can be endowed with a flip or rather the identity operators on $\ket{s^2=1/2}$ state, the corner transformation $\hat{U}_c$ with $\sigma_1$, $\sigma_2$ or the identity operators on the $\ket{s^2=1}$ state. However, for the choice of orientation we made in Figure $\ref{fig:start}$, $\hat{U}$ transformation does not affect the orientation of the cubies. As an example, we bring the $\hat{F}$ operator, which does affect the orientation of the cubies involved in the transformation. In order to build the $F$ rotation on the front face, one may refer to Figure $\ref{fig:positions}$ for the translations and to Figure $\ref{fig:start}$ for the orientations (the red face is the front face), so the overall $\hat{F}$ operator can be written as
\begin{equation} \label{eq:Fe}
    \hat{F}_e = \hat{P}^e_{\sigma(a,b,c,d)} e^{i\frac{2\pi}{l}(y_a+x_a+y_b-x_b-x_c-y_c-y_d+x_d)}\hat{\sigma}_x^a \hat{\sigma}_x^b \hat{\sigma}_x^c \hat{\sigma}_x^d
\end{equation}
\begin{equation} \label{eq:Fc}
    \hat{F}_c = \hat{P}^c_{\sigma(a,b,c,d)} e^{i\frac{2\pi}{l}(y_a-x_b-y_c+x_d)}\hat{\sigma}_{x_A}^a \hat{\sigma}_{x_C}^b \hat{\sigma}_{x_A}^c \hat{\sigma}_{x_C}^d
\end{equation}
For all the cubies not involved in the $\hat{F}$ transformation, an identity operator is applied, and their state is not altered. Referring to Figure $\ref{fig:grid}$, instead of $a$, $b$, $c$, $d$ there would be $1$, $9$, $5$, $11$ for the edges, $16$, $18$, $17$, $13$ for the corners. In Section \ref{sec:A}, a list of all the generators of the Rubik's group is provided.

\subsection{Rubik's group is unitary}

The spin operator $\sigma_x$ acts on a $\mathbb{R}^2$ space, while the spin operators $\sigma_{x_A}$ and $\sigma_{x_C}$ on a $\mathbb{R}^3$ space (corners), and in turn translations $\hat{K}_i$ on $L^2[0;l]$, respectively. Two operators acting on different spaces commute each other, which makes commute spin operators and translations, as required in equation ($\ref{eq:commuting_complete}$). However, permutations make the Rubik's group not abelian. The adjoint of $\hat{F}$ in Eq. ($\ref{eq:Fe}$) is
\begin{equation}
    \hat{F}_e^\dagger = e^{-i\frac{2\pi}{l}(y_a+x_a+y_b-x_b-x_c-y_c-y_d+x_d)}\hat{\sigma}_x^a \hat{\sigma}_x^b \hat{\sigma}_x^c \hat{\sigma}_x^d [\hat{P}^e_{\sigma(a,b,c,d)}]^\dagger
\end{equation}
Contracting $\hat{F}_e$ with $\hat{F}_e^\dagger$, the result is
\begin{equation}
    \hat{F}_e \hat{F}_e^\dagger = \hat{P}^e_{\sigma(a,b,c,d)} [\hat{P}^e_{\sigma(a,b,c,d)}]^\dagger = \hat{P}^e_{\sigma(a,b,c,d)} \hat{P}^e_{\sigma^{-1}(a,b,c,d)} = \hat{\mathbb{I}}_e
\end{equation}
An analogous result comes for $\hat{F}_c$ and $\hat{F}_c^\dagger$. The overall result is $\hat{F} \hat{F}^\dagger=\hat{\mathbb{I}}$, i.e. the identity acting on the space of the Cube.
The same result holds for any generator of the Rubik's group in equation ($\ref{eq:Rubikgroup}$), as far as it can be generalized for any $(a,b,c,d)$ tuple of cubies. 

\section{Strategy and results}
\label{sec:results}

We now turn our attention to the exploitation of the quantum formalism of the Cube by building an environment and 
reward functions so to implement a deep reinforcement learning agent aimed to solve it. 

\subsection{Reinforcement Learning}

\begin{figure}
\subfloat[][]{\includegraphics[width=.50\textwidth]{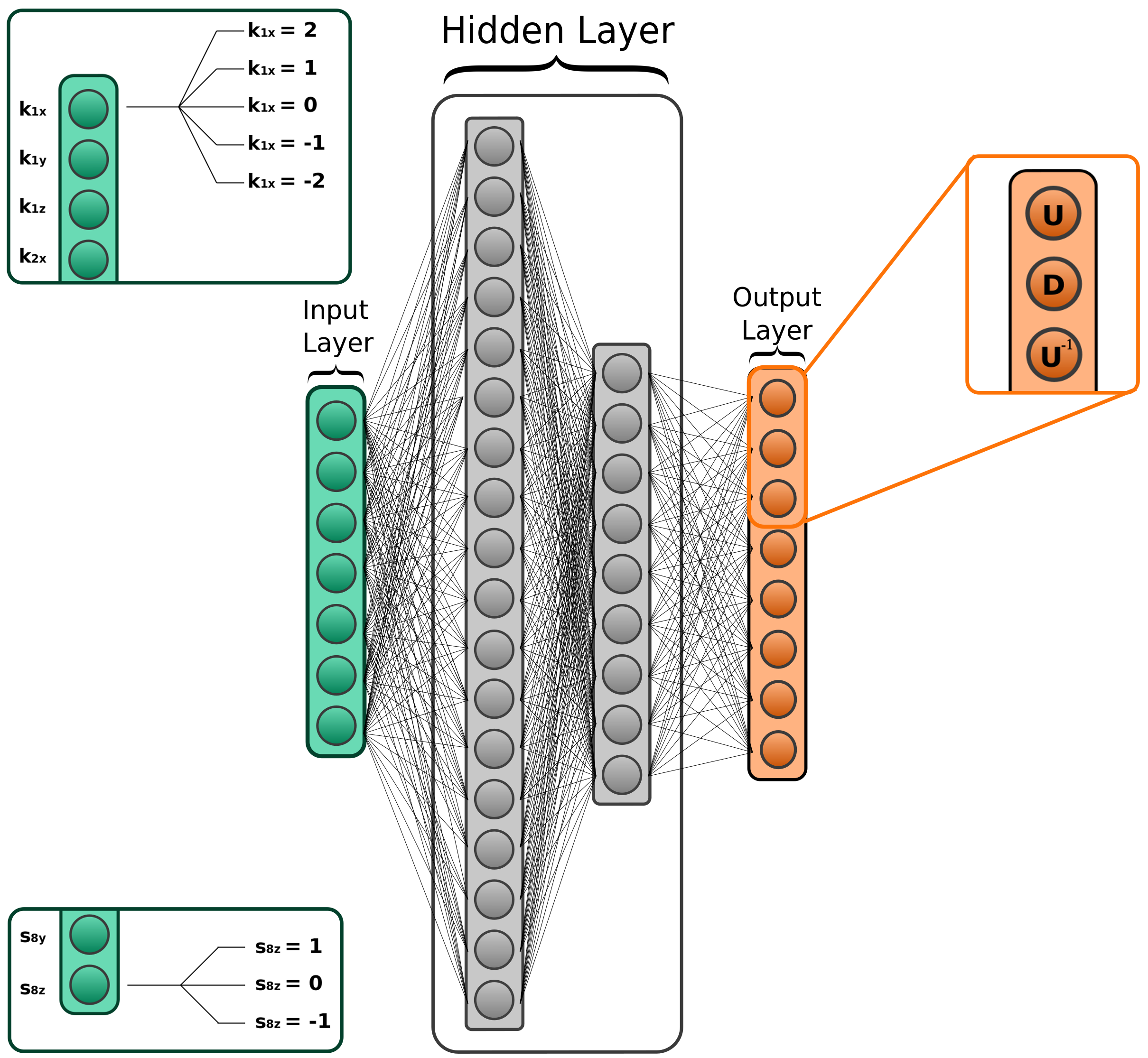}}
\subfloat[][]{\includegraphics[width=.50\textwidth]{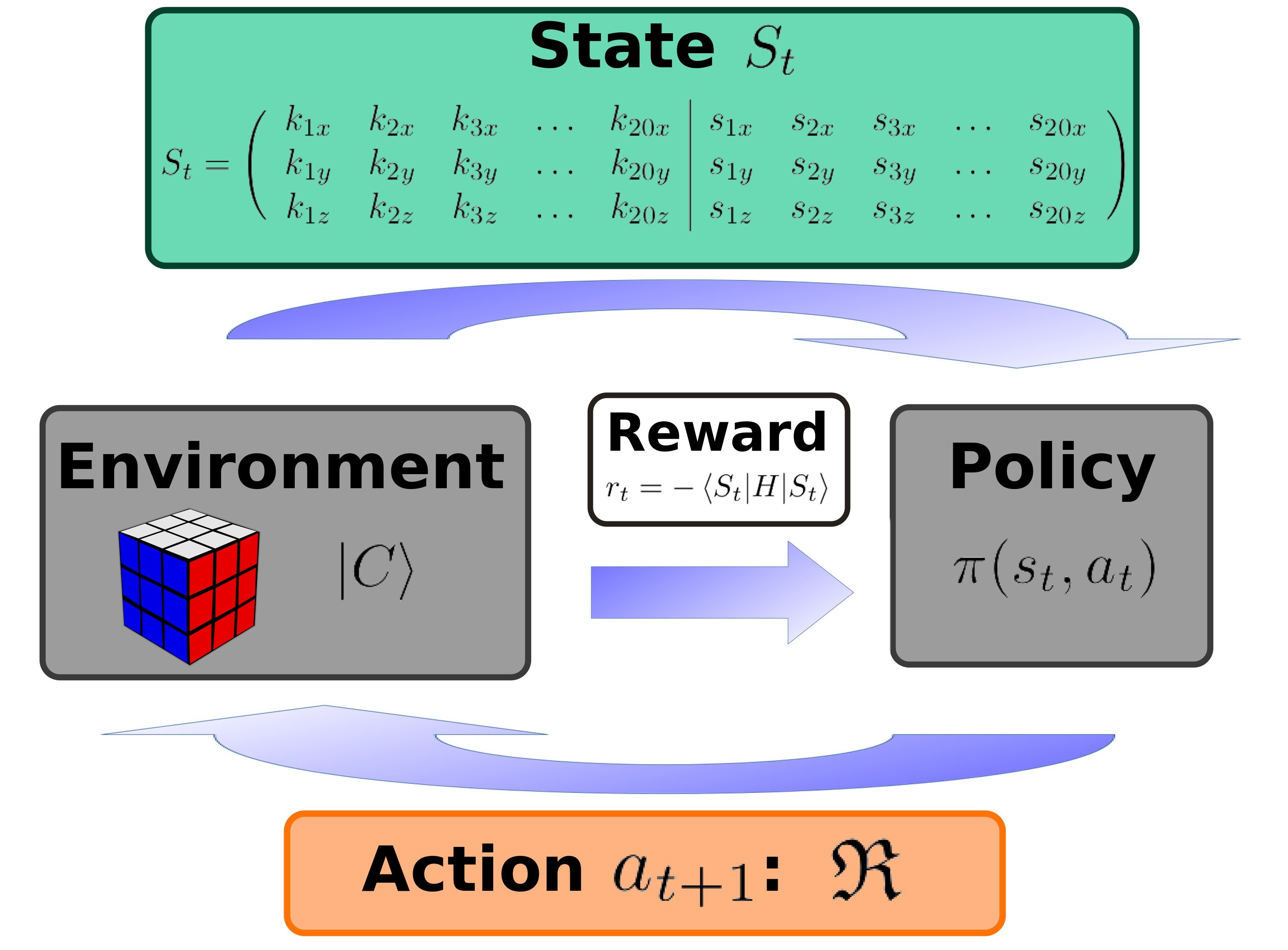}}\\ 
\centering\caption{(a) A scheme of the neural network architectures deployed during the training of the agent in the four phases. The frames on the left represent the possible values the input neuron can assume, for example for the edge momenta (above) and the corner spin (bottom). For each phase of the training, the input neurons assume a precise spectrum of values. On the right, the output neurons assume the label of a precise action the agent is able to take. (b) Scheme of interaction between environment and agent in Reinforcement Learning. Once a state and a reward are given, the agent will improve its policy and take a new action (from the Rubik's group $\mathfrak{R}$) on the environment.}
\label{fig:settings}
\end{figure}

Reinforcement Learning is a branch of Machine Learning which focuses on developing a strategy to solve a particular class of problems~\cite{sutton}. In these algorithms, it is given an environment that provides a set of possible states $\{s_i\}_{i=1}^N$ (the inputs) and possible actions $\{a_i\}_{i=1}^M$ (the outputs). When an action $a$ is taken, the current state $s$ of the environment is transformed into a new one $s'$. The purpose of the learning consists of training an artificial intelligence agent in order to define the optimal policy, i.e. which actions lead to the solution of the problem, represented by a target state $s_0$. Once trained, such policy predicts the best sequence of actions from a given starting state to achieve the goal state $s_0$. To evaluate the effectiveness of such policy, it is used a scalar reward $r$ returned to the agent after each couple action-state is sampled. The agent aims to maximize the total discounted return, given by the sum of all the rewards gained at the end of the problem. A scheme for a Reinforcement Learning algorithm is given in Figure $\ref{fig:settings}$. In Reinforcement Learning, setting a good reward $r(a, s)$ is the key feature, as it allows the agent to learn which are the best policies.


\subsection{The Hamiltonian approach}

Introducing and manipulating Hamiltonian rewards could be useful when dealing with physical systems in Reinforcement Learning. In the first place, the Hamiltonian gives a hint about how to build a $r : \mathcal{A} \times \mathcal{S} \to \mathbb{R}$ function, which is the core for any RL algorithm, setting the ground state of the Hamiltonian as the goal state of the problem. If such goal state is unique, the ground state is required to have no degeneracy. Secondly, by the usage of a Hamiltonian operator, Reinforcement Learning can be endowed with the mature field of minimization techniques.
In fact, when dealing with a RL problem, the training of the agent is very susceptible to the reward it is given. Introducing a Hamiltonian as reward, during the training it is possible to test and improve the performances of the agent by tuning some coefficients (for instance, the $J^{ij}$ and $B^{il}$ coefficients for the Ising model, see the equations below).

As said, a Hamiltonian is used to define the energy metric. To teach the agent how to reach the target state $\ket{s_0}$, we have introduced the following reward:
\begin{equation}
    r(s,a) = - \braket{s|\hat{U}^\dagger(a) \hat{H} \hat{U}(a)|s} = - \braket{s'|\hat{H}|s'}
\end{equation}
calling $\hat{U}(a)$ the unitary transformation implemented by $a$, while $\ket{s}$ is a generic state and $\ket{s'}$ the new state after the $\hat{U}(a)$ transformation. The reward the agent will gain is always negative, except when it reaches the ground state which returns a null reward. A boosted premium (i.e. a positive reward value) was added when the agent reaches the goal state, in order to increase the prediction of discounted return towards the ground state. The algorithm stores the $r_t$ reward as follows:
\begin{equation} \label{eq:reward}
    s_t \overset{a_t}{\longrightarrow} s_{t+1} \Rightarrow r_t = r(s_t,a_t) = - \braket{s_t|\hat{U}^\dagger(a_t) \hat{H} \hat{U}(a_t)|s_t} = - \braket{s_{t+1}| \hat{H} |s_{t+1}}
\end{equation}
where $s_t$ is the state of the Cube at the $t$-th step, $a_t$ the action taken from such a state, and $s_{t+1}$ is the state following $s_t$ by the action $a_t$.

In Section $\ref{sec:CubConf}$, operators such as momentum $\hat{\textbf{K}}$ or spin $\hat{\textbf{S}}$ were introduced to build a unitary representation for the Rubik's group. These observables can be employed to build Hamiltonians for the Rubik's Cube. As already remarked, dynamics does not furnish any equation to build the state of the Cube, as it happens when solving Schrodinger's equation. However, dynamics can be exploited to provide other information, for example how far the system is from its solved configuration.

The dynamics of the Cube is set on a grid, on whose sites are defined all the values of spin and momentum for each cubie. The quantum Ising model displays a Hamiltonian which suits quite well such a dynamics:
\begin{equation} \label{eq:isingmodel}
    \hat{H} = - \sum_{i,j} J_{ij} \hat{\sigma}_{i}^z \hat{\sigma}_{j}^z +  g \sum_i \hat{\sigma}_{i}^x
\end{equation}
When the interaction holds only between the nearest neighbours, $J_{ij} = J( \delta_{j,i+1}+\delta_{j,i-1})/2$, and  $J>0$ and $g>0$ are the exchange and the coupling constants~\cite{sachdev}, while $i$ and $j$ indices run along the number of spins on the grid. In the nearest-neighbours interaction, a parallel orientation of the $i$-th and $i+1$-th spin lowers the energy of the system, on the contrary, an anti-parallel aligning raises it. Instead, in another model, the so called spin glasses, the coupling constant $g$ is zero, the interaction $J_{ij}$ holds between all the spins and follows a Gaussian random distribution of probability~\cite{dutta2015quantum}. For our purpose, to describe the state of every cubie, two indices are required, the first pointing to the specific cubie, the other one labelling the component of a vector. For instance, the tensor $\hat{K}_{il}, \; i=1,..,8, \; l=1,2,3$ describes the $l$-th component of the $\hat{\textbf{K}}$ momentum operator for the $i$-th corner.


As previously said, introducing a Hamiltonian for the Cube corresponds to induce a distance in the phase space, i.e. to measure how far the cubies are from their solved configuration to a generic state $\ket{C}$. Such a distance should not be confused with the euclidean metric of a normed space. 
To introduce such distance for the position of the corner cubies, the previous models can be reassembled as follows (using Einstein notation):
\begin{equation} \label{eq:Hcorners}
    \hat{H}^c_{position} = J^{ij} \delta^{lm} [\hat{K}_{il}]^2 [\hat{K}_{jm}]^2 + B^{il} [\hat{K}_{il}]^2
\end{equation}
with $i,j$ indices running from $1$ to $8$ (the number of corners), the $m,l$ from $1$ to $3$ (the number of component of the spatial vector). Let $J^{ij}$ and $B^{il}$ be positive for any $i,j,l,m$, so that any Hamiltonian eigenvalue $E\geq 0$. Moreover, the state $E_0=0$, which fits the case $\hat{K}_{ij}\ket{C}=0 \; \forall i,j$, $\ket{C}$ being the overall state of the Cube, occurs only when every corner is located in its solved position. The Hamiltonian in equation ($\ref{eq:Hcorners}$) can be adapted as well for the position of the edges. In this case, the $i,j$ indices will run from $1$ to $12$, which is the total number of edges in the Cube, while $l$,$m$ still run from $1$ to $3$. The expression that such Hamiltonian assumes is:
%
\begin{equation} \label{eq:Hedges}
    \hat{H}^e_{position} = J^{ij} \delta^{lm} [\hat{K}_{il}]^2 [\hat{K}_{jm}]^2 + B^{il} [\hat{K}_{il}]^2
\end{equation}
Swapping the $\hat{\textbf{K}}_{i}$ operators with the $\hat{S}=\hat{\sigma}_{z}^c$ in section $\ref{sec:orientation}$, we obtain a Hamiltonian for the corner spins:
\begin{equation} \label{eq:Hspincorners}
    \hat{H}^c_{spin} = J^{ij} [\hat{S}_i]^2 [\hat{S}_j]^2 + B^i [\hat{S}_{i}]^2
\end{equation}
The $\hat{S}$ operator is a scalar (while $\hat{\textbf{K}}$ is a vector), and thus carry only one index, running along the number of cubies. For the edge spins indeed, the spectrum of the single $\hat{S}$ operator needs to be shifted, which can be accomplished defining the operator $\hat{S'}= \hat{\sigma}_z^e-1/2$. With such a shift, when an edge is oriented the corresponding spin operators $\hat{S}'$ returns a $0$ eigenvalue, while when disoriented a $-1$ one. Inserting $\hat{S'}_i$ operators in equation ($\ref{eq:Hspincorners}$), the Hamiltonian for the edges spins is available:
\begin{equation} \label{eq:Hspinedges}
    \hat{H}^e_{spin} = J^{ij} [\hat{S'}_i]^2 [\hat{S'}_j]^2 + B^i [\hat{S'}_i]^2
\end{equation}
As the eigenvalues for the $[\hat{S'}_i]^2$ are $1$ and $0$, the ground state for $\hat{H}^e_{spin}$ corresponds to $E_0=0$.

\subsection{Phases of the game: the QUBE algorithm}
\label{sec:4phases}

The solution of the Cube is reached when all of the mentioned Hamiltonians reach simultaneously their ground state. It is possible to formulate a global Hamiltonian $\hat{H}$ as their sum:
\begin{equation} \label{eq:Hamiltonians}
    \hat{H} = \hat{H}^e_{spin}+\hat{H}^c_{spin} + \hat{H}^e_{position} + \hat{H}^c_{position}
\end{equation}
As far as any Hamiltonian has strictly positive eigenvalues, except for the ground state (for which the energy value is zero), the ground state $\ket{GS}$ for the $\hat{H}$ Hamiltonian in equation ($\ref{eq:Hamiltonians}$) corresponds to the null eigenvalue:
\begin{equation} \label{eq:HamiltoniansGS}
    \hat{H} \ket{GS} = \left( \hat{H}^e_{spin}+\hat{H}^c_{spin} + \hat{H}^e_{position} + \hat{H}^c_{position} \right) \ket{GS} = 0
\end{equation}
When the state of the Cube matches the ground state, as in equation ($\ref{eq:HamiltoniansGS}$), the cubies (both corners and edges) are correctly oriented and positioned, which means the solved state has been reached. The four Hamiltonians have degeneracy on their ground states, for example there exist many states with corners and edges out of their solved position but still oriented in the correct way (referring to Figure $\ref{fig:start}$, you may rotate the upper layer without disorienting the Cube). On the other hand, $\ket{GS}$ is the sole state of the Cube with null eigenvalues for all of the four Hamiltonians, thus the $\hat{H}$ Hamiltonian has no degeneracy on its ground state. However, training an agent to reach the ground state via the global Hamiltonian $\ket{H}$ has proven to be a hard task. In the first place, there is an excess of data to store in memory, which can slow the learning process or even obstruct it. Collecting a huge amount of input parameters prompts to install a larger network architecture, whose training requires a higher computational time.
In second place, the further the system is from the solved state, the higher the eigenvalues of the Hamiltonian should be, but this condition does not hold necessarily. For instance, states closer to the solution may have a higher energy than other states further from it. When dealing with a huge number of states, it is might-impossible to define a Hamiltonian whose spectrum suits perfectly the distance from any state to the solution. In fact, it may happens to have a Hamiltonian spectrum with a high number of local minima, i.e. some states, reachable by a smaller number of moves, have higher energy than other states available by a further number of moves. Such a situation should be avoided, as it make the agent to forget the strategy it developed. The risk is that it tends to aim these minima instead of the solution. Decreasing the complexity of the spectrum may help to reduce the presence of local minima, and makes the learning process more linear.
The choice consists of optimizing each Hamiltonian separately by a sequence of four phases. Once the ground state of the four Hamiltonians is reached, the Cube is solved. Every phase corresponds to a specific Hamiltonian from equation ($\ref{eq:Hamiltonians}$) to optimize.
This phase-by-phase approach resembles the layer method or Kociemba's algorithm, but instead of focusing on the the colour of the faces, the order is based on momentum and spin of the cubies. In order to prevent that the moves of the next phase undo the current state of the Cube, the twelve fundamental moves from Rubik's group do not suite this purpose. A phase is solved when the state of the Cube matches with the ground state of one of the Hamiltonians in equation ($\ref{eq:Hamiltonians}$). Each phase must deploy a set of moves able to find the ground state of the current Hamiltonian, without affecting the ground state of the previously optimized ones.
When one of these Hamiltonians lies in its ground state, a symmetry is preserved. For instance, when the state $\ket{s}$ of the Cube is such that $H^c_{spin}\ket{s}=0$ and $H^e_{spin}\ket{s}=0$, the spins of all the cubies are correctly aligned. We want then to find some sets of moves capable of preserving such symmetries. For example, if the $\hat{H}_i$ Hamiltonian lies in its ground state $\ket{s_j}$, we should be able to find a set $\{ \hat{A}_{i+1} \}$ of transformations such that
\begin{equation} \label{eq:symmetry}
\hat{H}_i \ket{s_j}=0, \quad \hat{H}_i \, \hat{A}_{i+1} \ket{s_j} = \hat{H}_i \ket{s_{j+1}} = 0 \quad \forall \hat{A}_{i+1} \in \{\hat{A}_{i+1} \}
\end{equation}
If the condition in equation ($\ref{eq:symmetry}$) holds not only for $\ket{s_{j+1}}$, but for any $\ket{s_{j+n}}$, $n$ being the number of $\hat{A}_{i+1}$ moves applied after $\ket{s_j}$, it means the symmetry is preserved by such a set of $\{ \hat{A}_{i+1} \}$ transformations. When the symmetry is preserved, all the $\ket{s_{j+n}}$ states still correspond to the ground state for the $\hat{H}_i$ Hamiltonian. The same condition in equation ($\ref{eq:symmetry}$) can be written as
\begin{equation}
    \left[\hat{H}_i;\hat{A}_{i+i} \right]\ket{s_j} = 0 \quad \forall \hat{A}_{i+1} \in \{\hat{A}_{i+1} \}
\end{equation}
The commutative relation above does not hold in general, but only if the $\ket{s_j}$ is the ground state for $\hat{H}_i$. To build an algorithm, i.e. to combine Hamiltonians and opportune transformations to reach the solution, we must study which moves preserve the symmetries of the Cube, and then assemble them to build an algorithm capable of reaching the solution. When all the symmetries of the Hamiltonians in equation ($\ref{eq:Hamiltonians}$) are preserved, the solution has been reached. The sets of moves introduced in the algorithm, and order by which they are implemented, are reported in Table $\ref{tab:QUBE}$, where the training of the agent is explained.
For the choice of orientation we have made in Figure $\ref{fig:start}$, in the Rubik's group $\mathfrak{R}$ only $U$, $D$ and their inverse elements $U^{-1}$, $D^{-1}$ preserve orientation, while all the other elements do not. However, double combinations such as $F^2$, $B^2$, $L^2$, $R^2$ do.
Another very frequent combination of generators is the so called commutator. A commutator $C$ is built by two moves, $A$ and $B$, then combining them by
\begin{equation} \label{eq:commBasic}
    C = A \circ B \circ A^{-1} \circ B^{-1}
\end{equation}
Commutators, for instance, can be used to orient corners, any non-oriented corner can be correctly oriented by applying two or four times an appropriate commutator. Indeed, a combination $C^2$ of two commutators can be applied as an action when orienting the corners. Furthermore, commutators can be in turn combined to rotate three edges without affecting the position of the other cubies. Let be $A$ and $B$ the rotation of two opposite faces, and $M$ the rotation of the face in between, it is possible to build two different couples of commutators:
\begin{equation}
    C_1 = A \circ M \circ A^{-1} \circ M^{-1}, \quad
    C_2 = B \circ M \circ B^{-1} \circ M^{-1}, \quad
\end{equation}
The rotation $C_3$ of three edges is accomplished by composing the commutators in the following way:
\begin{equation} \label{eq:compositioncomm1}
    C_3 = C_1 \circ C_2 \circ C_1^5 \circ C_2^5
\end{equation}
For the inverse rotation, it suffices to swap $C_1$ and $C_2$:
\begin{equation} \label{eq:compositioncomm2}
    C_4 = C_2 \circ C_1 \circ C_2^5 \circ C_1^5
\end{equation}
The algorithm we built assembles all these sets of moves, in order to find the minimum of the Hamiltonians via the Reinforcement Learning approach. We called this algorithm the QUBE algorithm. Its phases, with the corresponding Hamiltonians and sets of moves, are reported in Table $\ref{tab:QUBE}$. Each phase of the game is implemented by a Deep Double Q-Learning algorithm (or DDQN, Deep Double Q-Network), which is a quite popular family of Reinforcement Learning algorithms. The overall QUBE algorithm collects the results from all of these four phases, in order to solve the Rubik's Cube.

\begin{table}
\begin{tabular}{ |p{.3cm}|p{3.5cm}||p{2.5cm}|p{4.5cm}|  }
 \hline
 \multicolumn{4}{|c|}{\textbf{QUBE Algorithm}} \\
 \hline
 \textbf{\#} & \textbf{Phase of the game} & \textbf{Hamiltonian} & \textbf{Set of moves} \\
 \hline
 1 & Orienting the edges   & $\hat{H}^e_{spin}$ & $\{U,U^{-1}, D,D^{-1}, B, B^{-1}, $ $ F,F^{-1}, L,L^{-1}, R,R^{-1}\}$
 \\
 \hline
 2 & Orienting the corners&   $\hat{H}^c_{spin}$  & $\{$commutators, $U,D \}$
 \\
 \hline
 3 & Positioning the corners & $\hat{H}^c_{position}$ & $\{U,U^{-1}, D,D^{-1}, B^2, F^2, $ $ L^2, R^2\}$
 \\
 \hline
 4 & Positioning the edges  & $\hat{H}^e_{position}$ & $\{C_3$ Eq. ($\ref{eq:compositioncomm1}$), $C_4$ Eq. ($\ref{eq:compositioncomm2})\}$\\
 \hline
\end{tabular}
\caption{\label{tab:QUBE} Training phases of the agent. The Hamiltonians, from equation ($\ref{eq:Hamiltonians}$), furnish the reward, as shown in equation ($\ref{eq:reward}$). The sets of available moves to solve each phase, and preserve the previous results, are reported as well. The purpose is to train each agent to find the ground state of each Hamiltonian, via the sets of moves given for each phase. Training the agent to solve one phase consists of developing a neural networks, able to store the necessary information to reach the ground state of the specific Hamiltonian via the corresponding set of actions. Once all the four neural networks are available, combining their actions it is possible to solve the Rubik's Cube. In the first phase, the agent uses the twelve fundamental moves from Rubik's group, to align the edge spins in their correct orientation. In the second phase the agent tries to align the corner spins, by using a set of commutators. In the third phase, the agent proceeds with allocating the corners in their correct position, via $U$, $D$, their inverse actions $U^{-1}$, $D^{-1}$ and the quadratic operators $B^2$, $F^2$, $L^2$, $R^2$, all of which do not alter the orientation of the cubies. At last, the final phase consists of allocate the edges in their correct position, via the combination of commutators reported in equations ($\ref{eq:compositioncomm1}$) and ($\ref{eq:compositioncomm2}$).}
\end{table}

\subsection{Results}

Each phase requires a specifically designed neural networks architecture. The input vector is given by the eigenvalues of the spins or momenta of the cubies, while the output vector is formed by all the possible actions the agent can take. All the NN have architectures involving two hidden layers, plus the input and output ones. To build the hidden layers, we took an empiric rule, which has proven to be quite efficient, i.e. to dispose of the second layer with $8.5$ times the neurons has the input layer, while the third layer disposes of half of the neurons of the previously hidden layer, ~\cite{california}.

\begin{figure}
\subfloat[][]{\includegraphics[width=.50\textwidth]{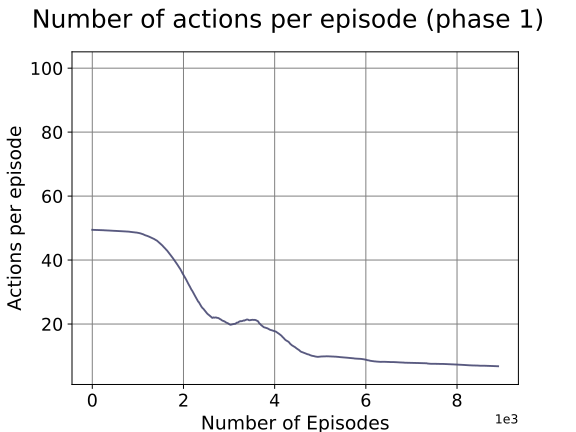}}
\subfloat[][]{\includegraphics[width=.50\textwidth]{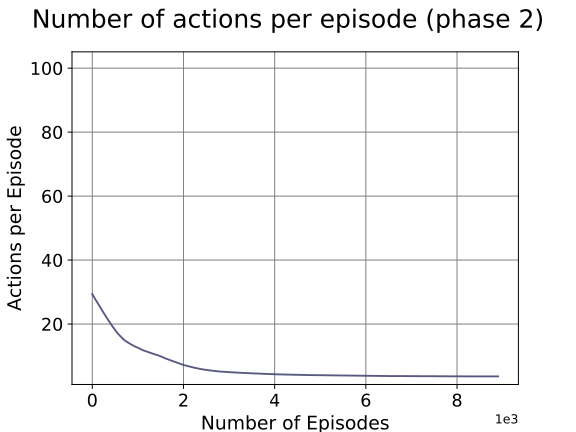}}\\
\subfloat[][]{\includegraphics[width=.50\textwidth]{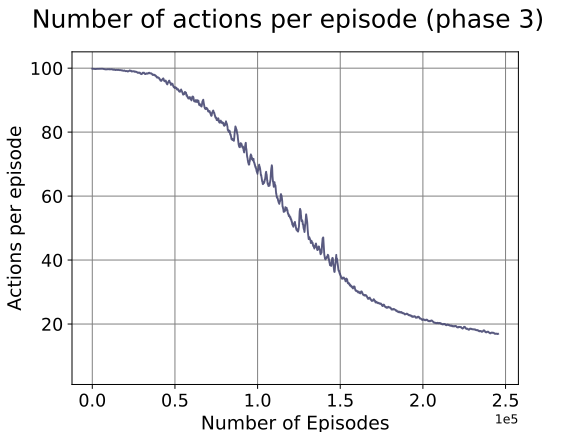}}
\subfloat[][]{\includegraphics[width=.50\textwidth]{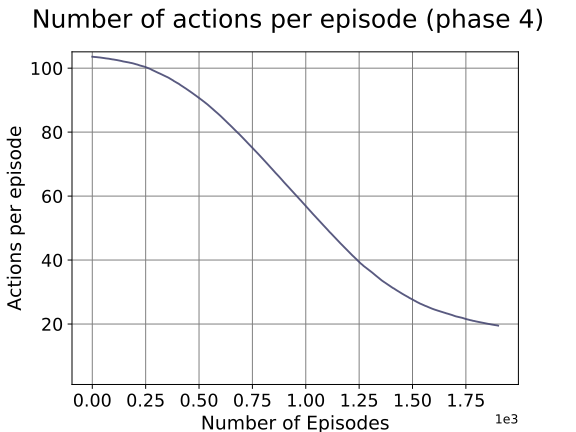}}\\ 
\centering\caption{In the four graphs above, it is reported the number of actions which the agent takes to solve an episode during the training. Learning to solve an episode, the trend of the agent is to take a smaller number of actions by the training. Orienting the cubies takes far less actions than positioning them correctly (a) Number of actions which the agent takes to orient the edges (phase 1). (b) Number of actions which the agent takes to orient the corners (phase 2). (c) Number of actions which the agent takes to position correctly the corners (phase 3). (d) Number of actions which the agent takes to position correctly the edges (phase 4). }
\label{fig:length}
\end{figure}

\begin{figure}
\subfloat{\includegraphics[width=.65\textwidth]{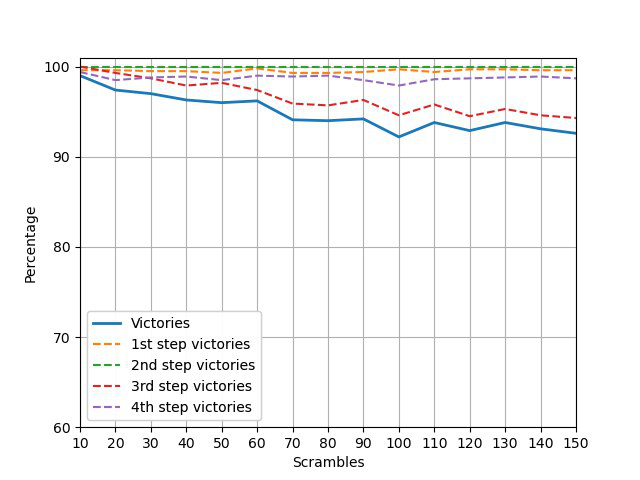}}\\
\centering\caption{Percentage of solutions by the number of scrambles. To test the phase of training, we stressed the agent by showing it $1000$ configurations, scrambled by the $12$ fundamental moves of the Rubik's group. The percentage of victories by all the four phases (solid line) are shown. In dashed lines, the percentage of victories for the first, the second, the third and the forth phases. Notice that the second phase runs into a number of defeats equal to zero. }
\label{fig:final}
\end{figure}

As long as each episode is solved in four phases, four NNs were trained, see Table $\ref{tab:QUBE}$, with the task to find the ground state of a peculiar Hamiltonian from those in Section $\ref{sec:4phases}$. To train the agent, a stage of exploration is followed by the exploitation of the achieved knowledge. First, the Cube is scrambled from the solved state, by a random sequence from the available actions, with a random length from $1$ to $50$ moves. Next, the agent starts taking some actions to reach the solution. At the beginning of the training, the agent chooses a random action with probability $1$, in order to explore the phase space for the action-state couples in $\mathcal{A}\times\mathcal{S}$. By proceeding with the moves, such probability decreases, and the agent is driven by a policy of maximum reward rather than a random choice. The rate of such decreasing is reported as a hyperparameter, called random action decay, for each of the four phases in Tables $\ref{tab:1}$, $\ref{tab:2}$, $\ref{tab:3}$ and $\ref{tab:4}$. The results from the test of the game are reported in Figure $\ref{fig:final}$. By increasing the number of scrambles when testing the training of the agent, the percentage of solutions does not move from a $90$\% value. In Figure $\ref{fig:final}$, it is possible to see the percentage of success from the four phases and the total percentage of success (all above 90\%). In the first and fourth phases, the asymptotic percentage of success per number of episodes is 99.5\%  and 98.8\% respectively, the complete $100$\% of solutions is reached in the second phase, while in the third such percentage fluctuates around $94.5$\%. During the third phase, the agent deals with a set of moves where a high number of cubies are involved at the end of the action: in the third phase, a total of $8$ cubies per move are involved, while in the forth just $3$ edges. The more cubies are involved, the more difficult will be to move them in the correct positions, and the more minima will occur in the energy landscape. For this reason, the training of the agent in the third phase takes far more epochs than in the other phases (see Figure $\ref{fig:length}$). In each phase, the agent was trained by showing it a Cube scrambled by $50$ moves. In the phase of test, while achieving results close to $100$\% of success when scrambling the Cube in a range of $50$ actions, the overall success is asymptotically stable above $94$\%.

Each test has been supported by a graphical environment, printing step by step the Cube during the process of scrambling and solving. All the successful tests match the graphical evidence to have reached the solution of the game.

\begin{figure}
\subfloat[][]{\includegraphics[width=.50\textwidth]{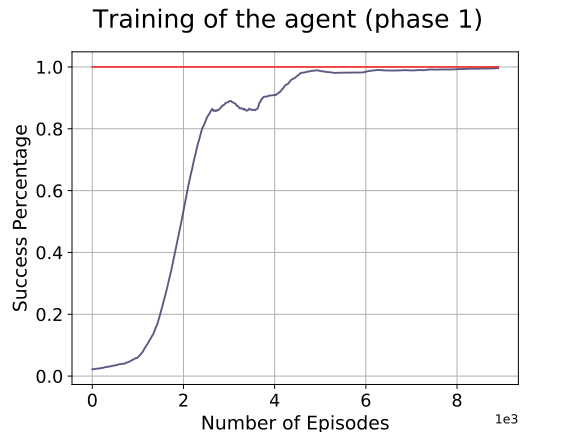}}
\subfloat[][]{\includegraphics[width=.50\textwidth]{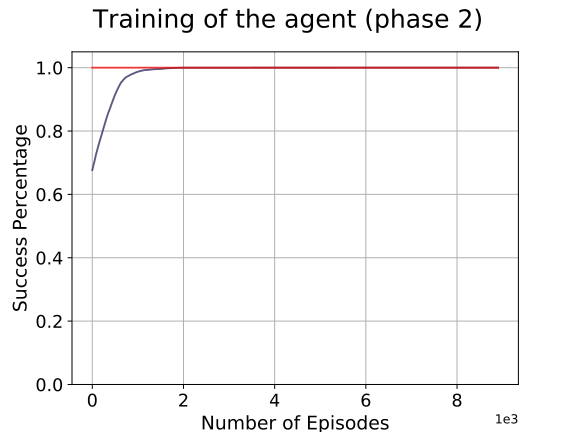}}\\
\subfloat[][]{\includegraphics[width=.50\textwidth]{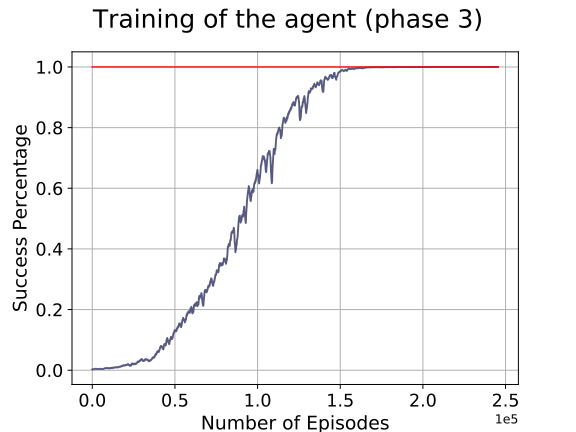}}
\subfloat[][]{\includegraphics[width=.50\textwidth]{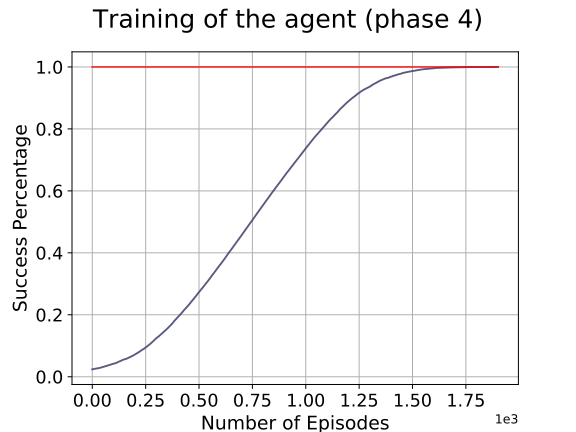}}\\ 
\centering\caption{Training of the agent in the four phases. On the x axis it is reported the number of training epochs of each phase, while on the y axis the percentage of success (from $0$, total failure, to $1$, complete success). The red line flags the 100\% of solutions. (a) The training of the agent to orient the edges took around $10^3$ episodes to reach an asymptotic $100$\% percentage of solving. (b) The training of the agent to orient the corners is very fast compared to the other phases, and very efficient, as the agent learns to solve $100$\% of cases. (c) The training of the agent to place correctly the corners is the longest of all, more than $10^5$ episodes of training. (d) The training of the agent to place correctly the edges. The number of episodes for the agent to learn is around $10^3$ episodes. }
\label{fig:training}
\end{figure}

\begin{table}
\begin{center}
\begin{tabular}{ |l|l|l| }
\hline
\multicolumn{3}{ |c| }{\textbf{DDQN ALGORITHM (phase 1)}} \\
\hline
\textbf{Area related} & \textbf{Hyper-parameter} & \textbf{Value} \\ \hline
\multirow{2}{*}{Architecture} 
 & \# input neurons (parameters) & 12 \\
 & \# hidden layer neurons & 100, 50 \\
 & \# output neurons (actions) & 12 \\
 & activations & ReLu, ReLu, Linear \\ \hline
\multirow{3}{*}{Learning} & optimizer & Adam \\
 & learning rate & 0.0001 \\
 & random action decay & 0.9995 \\
 & final premium of score & +5000 \\ \hline
\multirow{3}{*}{Algorithm} & target update frequency & 100 episodes \\
    & batch size & 1240 \\
 & memory size & $10^4$ experiences \\
 & max steps per episode & \# scrambles + 5 \\
 \hline
 \end{tabular}
 \caption{\label{tab:1} List of the hyper-parameters and their values used by DDQN agent in orienting the edges.}
 \end{center}
 \end{table}
 
\begin{table}
\begin{center}
\begin{tabular}{ |l|l|l| }
\hline
\multicolumn{3}{ |c| }{\textbf{DDQN ALGORITHM (phase 2)}} \\
\hline
\textbf{Area related} & \textbf{Hyper-parameter} & \textbf{Value} \\ \hline
\multirow{2}{*}{Architecture} 
 & \# input neurons (parameters) & 4 \\
 & \# hidden layer neurons & 35, 16 \\
 & \# output neurons (actions) & 3 \\
 & activations & ReLu, ReLu, Linear \\ \hline
\multirow{3}{*}{Learning} & optimizer & Adam \\
 & learning rate & 0.0001 \\
 & random action decay & 0.9995 \\
 & final premium of score & +5000 \\\hline
\multirow{3}{*}{Algorithm} & target update frequency & 100 episodes \\
    & batch size & 1240 \\
 & memory size & $10^4$ experiences \\
 & max steps per episode & \# scrambles + 5 \\
 \hline
 \end{tabular}
 \caption{\label{tab:2} Hyper-parameters and their values used by DDQN agent in orienting the corners.}
 \end{center}
 \end{table}
 
 \begin{table}
\begin{center}
\begin{tabular}{ |l|l|l| }
\hline
\multicolumn{3}{ |c| }{\textbf{DDQN ALGORITHM (phase 3)}} \\
\hline
\textbf{Area related} & \textbf{Hyper-parameter} & \textbf{Value} \\ \hline
\multirow{2}{*}{Architecture} 
 & \# input neurons (parameters) & 24 \\
 & \# hidden layer neurons & 200, 100 \\
 & \# output neurons (actions) & 8 \\
 & activations & ReLu, ReLu, Linear \\ \hline
\multirow{3}{*}{Learning} & optimizer & Adam \\
 & learning rate & 0.0001 \\
 & random action decay & 0.999995 \\
 & final premium of score & +5000 \\ \hline
\multirow{3}{*}{Algorithm} & target update frequency & 100 episodes \\
    & batch size & 1240 \\
 & memory size & $10^4$ experiences \\
 & max steps per episode & \# scrambles * 2 \\
 \hline
 \end{tabular}
 \caption{\label{tab:3} Hyper-parameters and their values used by DDQN agent in positioning the corners.}
 \end{center}
 \end{table}
 
 \begin{table}
\begin{center}
\begin{tabular}{ |l|l|l| }
\hline
\multicolumn{3}{ |c| }{\textbf{DDQN ALGORITHM (phase 4)}} \\
\hline
\textbf{Area related} & \textbf{Hyper-parameter} & \textbf{Value} \\ \hline
\multirow{2}{*}{Architecture} 
 & \# input neurons (parameters) & 36 \\
 & \# hidden layer neurons & 310, 115 \\
 & \# output neurons (actions) & 36 \\
 & activations & ReLu, ReLu, Linear \\ \hline
\multirow{3}{*}{Learning} & optimizer & Adam \\
 & learning rate & 0.0001 \\
 & random action decay & 0.9995 \\
 & final premium of score & +5000 \\ \hline
\multirow{3}{*}{Algorithm} & target update frequency & 100 episodes \\
    & batch size & 1240 \\
 & memory size & $10^4$ experiences \\
 & max steps per episode & \# scrambles + 5 \\
 \hline
 \end{tabular}
 \caption{\label{tab:4} Hyper-parameters and their values used by DDQN agent in positioning the edges.}
 \end{center}
 \end{table}

\section{Conclusions}
\label{sec:Conclusions}

The Rubik's Cube is described by the formalism of quantum mechanics. The Rubik's group can be mapped into a unitary representation.

The distance between states is based on the spectrum of Ising Hamiltonians, to estimate how far the actual state of the Cube is from its solved configuration. Such a distance is exploited to define a novel kind of Hamiltonian reward function in a Deep Reinforcement Learning environment. Four agents acting in sequence are trained according to their respective Hamiltonian based reward. 
The agents discover a path from any state in the phase space back to the solution by using only Hamiltonian based rewards. 

We have shown that it is possible to employ Deep Reinforcement Learning algorithms to exploit the symmetries of the system and solve a difficult combinatorial problem by using only Hamiltonian based reward functions. Such a physical approach may support new Reinforcement Learning algorithms, and link this branch of Artificial Intelligence to the new realm of adiabatic quantum computing which naturally fits the Ising model use to build the Hamiltonian rewards.

\section{Appendix A: generators of the Rubik's group in a unitary representation}
\label{sec:A}

In this section, we present a list of all the generators of the Rubik's group, which were listed as $\{U, D, F, B, L, R \}$ in Singmaster's notation. The labels of the cubies are reported in Figure $\ref{fig:grid}$, while the axes in Figure $\ref{fig:start}$. All the operators can be built as in Section $\ref{sec:action}$, by drawing a scheme as in Figure $\ref{fig:positions}$ for each operator. Here are reported all the operators which turn the layers of the Cube clockwise. To build the anti-clockwise operators, it is possible to repeat the inverse scheme, or to compute the adjoint of the clockwise operators.

\subsection{Generators preserving orientation}

By the choice of orientation we made in Figure $\ref{fig:start}$, $U$ and $D$ operators do not affect the global orientation of the Cube, but just permute the cubies. The labels of the cubies are the numbers in Figure $\ref{fig:grid}$.
\begin{equation}
    \hat{D}_e = \hat{P}_{\sigma(1,2,3,4)} e^{i\frac{2\pi}{l}(-x_1-y_1+x_2-y_2+x_3+y_3-x_4+y_4)}
\end{equation}
\begin{equation}
    \hat{D}_c = \hat{P}_{\sigma(13,14,15,16)} e^{i\frac{2\pi}{l}(-y_{13} + x_{14}+y_{15}-x_{16})}
\end{equation}
\begin{equation}
    \hat{U}_e = \hat{P}_{\sigma(5,6,7,8)} e^{i\frac{2\pi}{l}(x_5-y_5-x_6-y_6-x_7+y_7+x_8+y_8)}
\end{equation}
\begin{equation} 
    \hat{U}_c = \hat{P}_{\sigma(17,18,19,20)} e^{i\frac{2\pi}{l}(x_{17}-y_{18}-x_{19}+y_{20})}
\end{equation}

\subsection{Generators preserving edge orientation}

By the choice of orientation we made in Figure $\ref{fig:start}$, $L$ and $R$ operators do not affect the orientation of the edge cubies, but permute the edges and disorient four of the corners.
\begin{equation}
    \hat{L}_e = \hat{P}_{\sigma(4,10,6,9)} e^{i\frac{2\pi}{l}(y_6-z_6-y_9-z_9-y_4+z_4+y_{10}+z_{10})}
\end{equation}
\begin{equation} 
    \hat{L}_c = \hat{P}_{\sigma(15,19,18,16)} e^{i\frac{2\pi}{l}(z_{15}-y_{16} -z_{18} + y_{19} )} \sigma_{x_A}^{15} \sigma_{x_C}^{16} \sigma_{x_A}^{18} \sigma_{x_C}^{19}
\end{equation}
\begin{equation}
    \hat{R}_e = \hat{P}_{\sigma(2,11,8,12)} e^{i\frac{2\pi}{l}(y_2+z_2-y_8-z_8-y_{11}+z_{11}+y_{12}-z_{12})}
\end{equation}
\begin{equation} 
    \hat{R}_c = \hat{P}_{\sigma(13,17,20,14)} e^{i\frac{2\pi}{l}(z_{13}+y_{14}-y_{17}-z_{20})} \sigma_{x_A}^{13} \sigma_{x_C}^{14} \sigma_{x_C}^{17} \sigma_{x_A}^{20}
\end{equation}

\subsection{Generators which do not preserve orientation}

$F$ and $B$ do not preserve orientation, neither for the edges 
\begin{equation}
    \hat{F}_e = \hat{P}_{\sigma(1,9,5,11)} e^{i\frac{2\pi}{l}(x_1+z_1-x_5-z_5-x_9+z_9+x_{11}-z_{11})} \sigma_x^1 \sigma_x^5 \sigma_x^9 \sigma_x^{11}
\end{equation}
\begin{equation}
    \hat{F}_c = \hat{P}_{\sigma(13,16,18,17)} e^{i\frac{2\pi}{l}( x_{13} + z_{16} - z_{17} - x_{18}  )} \sigma_{x_C}^{13} \sigma_{x_A}^{16} \sigma_{x_A}^{17} \sigma_{x_C}^{18}
\end{equation}
\begin{equation}
    \hat{B}_e = \hat{P}_{\sigma(3,12,7,10)} e^{i\frac{2\pi}{l}( -x_3 + z_3 + x_7 - z_7 - x_{10} - z_{10} + x_{12} + z_{12} )} \sigma_x^3 \sigma_x^7 \sigma_x^{10} \sigma_x^{12}
\end{equation}
\begin{equation}
    \hat{B}_c = \hat{P}_{\sigma(14,20,19,15)} e^{i\frac{2\pi}{l}( z_{14} - x_{15} - z_{19} + x_{20} )} \sigma_{x_A}^{14} \sigma_{x_C}^{15} \sigma_{x_A}^{19} \sigma_{x_C}^{20}
\end{equation}

\bibliography{references}

\end{document}